\documentclass[%
 reprint,
superscriptaddress,
 amsmath,amssymb,
 aps,
floatfix,
]{revtex4-2}

\usepackage{graphicx}
\usepackage{xcolor}
\usepackage{lineno}
\usepackage{makecell}
\usepackage{booktabs}
\usepackage{soul}
\usepackage{siunitx}

\begin{document}

\title{Three-Wave Mixing as a High-Throughput Diagnostic Tool for Nonlinear Waveguides}

\author{Yi-Yang Liu}
\affiliation{Physics and Informatics Laboratories, NTT Research, Inc., Sunnyvale, CA 94085, USA}
\affiliation{School of Applied and Engineering Physics, Cornell University, Ithaca, NY 14853, USA}

\author{Kevin Chen}
\affiliation{Physics and Informatics Laboratories, NTT Research, Inc., Sunnyvale, CA 94085, USA}

\author{Kyungduk Kim}
\affiliation{Physics and Informatics Laboratories, NTT Research, Inc., Sunnyvale, CA 94085, USA}

\author{Ryotatsu Yanagimoto}
\affiliation{Physics and Informatics Laboratories, NTT Research, Inc., Sunnyvale, CA 94085, USA}
\affiliation{School of Applied and Engineering Physics, Cornell University, Ithaca, NY 14853, USA}

\author{Edwin Ng}
\affiliation{Physics and Informatics Laboratories, NTT Research, Inc., Sunnyvale, CA 94085, USA}

\author{Timothy P. McKenna}
\affiliation{Physics and Informatics Laboratories, NTT Research, Inc., Sunnyvale, CA 94085, USA}

\author{Marc Jankowski}
\email{marc.jankowski@ntt-research.com}
\affiliation{Physics and Informatics Laboratories, NTT Research, Inc., Sunnyvale, CA 94085, USA}

\date{\today}

\begin{abstract}
The recent development of wavelength-scale nonlinear waveguides highlights the need for new diagnostic methods that characterize both waveguide dispersion and fabrication-induced spatial inhomogeneities. While tightly confining waveguides enable compact and highly efficient nonlinear devices, nanometer-scale waveguide nonuniformities can limit the effective interaction length and conversion efficiency. In this work, we establish a high-throughput diagnostic method for second-order nonlinear waveguides based on near-degenerate three-wave mixing, hereafter referred to as sum-frequency tomography (SFT). In contrast with conventional diagnostics based on second-harmonic generation, SFT produces two-dimensional transfer functions that allow dispersion orders to be characterized separately from longitudinal phase-mismatch variations. We validate this technique using commercially available PPLN waveguides and show that SFT recovers the group-velocity mismatch between the interacting waves, the group-velocity dispersion of the fundamental, and a low-order polynomial approximation of the longitudinal phase-mismatch profile. SFT avoids the experimental complexities of phase-sensitive or spatially resolved measurements, enabling the characterization of individual waveguides in seconds.
\end{abstract}

\maketitle
\section{Introduction}
Nonlinear optics leverages interactions between light waves to enable functions that are impossible with purely linear optics~\cite{Boyd_NLO_2008}. Devices based on third-order nonlinearities have advanced rapidly through the integration of nonlinear materials with optical waveguides, which provide tight transverse field confinement without the penalties of diffraction or spatial walk-off~\cite{agrawal2012nonlinear}. Control over waveguide geometry also enables dispersion engineering, allowing access to qualitatively new nonlinear operating regimes through the interplay between dispersion and nonlinearity~\cite{hasegawa1973transmission-soliton,tai1986observation-MI, solli2007optical-rogue-wave,grelu2012dissipative}. Second-order ($\chi^{(2)}$) nonlinear optics is now undergoing a similar transformation. The nonlinear interaction strength in $\chi^{(2)}$ waveguides is sufficiently large to enable near-unity conversion efficiency in single-pass, continuous-wave-pumped devices~\cite{Pelc2011}. Packaged modules based on these technologies have turned what was once accessible only in academic labs into commercially available devices, with applications including broadband, low-noise amplifiers~\cite{kobayashi2021wide,shimizu2022ppln,shimizu2023effect,shimizu2024and,umeki2011phase,asobe2012phase,umeki2013line,umeki2014first,kazama2016low,olsson2018long,kazama2021over}, optical signal processing~\cite{kawanishi2001high,kurz2001optical,langrock2006alloptical,jiang2005four,tu2023photonic,jin2025thin,song2025integrated}, and optical computing~\cite{McMahon2016}. The parametric interactions mediated by the $\chi^{(2)}$ nonlinearity allow the generation, manipulation, and detection of quantum light and have played a central role in the development of quantum photonic technologies~\cite{xin2022spectrally, kashiwazaki2021fabrication, nehra2022few, karnik202618, 
park2024single, 
hu2026efficient-adaptive-PM}. State-of-the-art $\chi^{(2)}$ nanowaveguides now combine dense integration and dispersion engineering with ultra-low-power operation. These devices have sufficient interaction strength to enable unity conversion efficiency with only thousands of pump photons~\cite{lu2020toward, zhao2022ingap, guo2022femtojoule, yanagimoto2022temporal}, opening pathways to a wide range of emergent applications in both the classical~\cite{li2023all-ReLu, fu2026passive-activation, dean2026low} and quantum~\cite{vendromin2026non-Gaussian, yanagimoto2023-QND,  yanagimoto2024mesoscopic} realms.

These advantages come at the cost of stringent fabrication requirements, particularly for devices designed for phase-matched nonlinear interactions. Since the geometry of the waveguide determines the propagation constant of the interacting waves, small deviations from the nominal geometry can introduce a residual phase mismatch. Moreover, longitudinal variations of the waveguide geometry can produce a spatially varying phase mismatch that limits the effective interaction length of the waveguide and therefore the attainable conversion efficiency. These effects can be mitigated in diffuse and machined waveguides by using noncritical designs~\cite{bortz1994noncritical}, where the phase mismatch becomes a weak function of the waveguide shape. The situation becomes far more severe in tightly confining nanophotonic waveguides. Here, the strong geometric dispersion precludes noncritical operation~\cite{Kuo2021}, and the phase mismatch induced by even nanometer-scale variations in waveguide geometry can degrade device performance~\cite{chen2024adapted,zhao2023unveiling}. Geometry errors on the scale of tens of nanometers can strongly alter the dispersion relations, further limiting the yield of more sophisticated dispersion-engineered waveguides. New diagnostic techniques are therefore needed to simultaneously characterize spatial inhomogeneities and waveguide dispersion. Furthermore, these diagnostics must offer high throughput to keep pace with the growing production volumes of modern nonlinear photonics.

\begin{figure*}[t]
    \centering
    \includegraphics[width=0.8\textwidth]{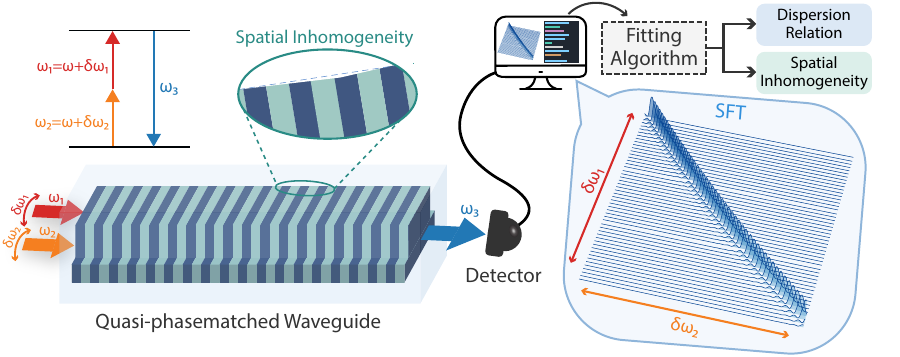}
    \caption{Conceptual overview of sum-frequency tomography. Two continuous-wave lasers with frequencies $\omega_1=\omega+\delta\omega_1$ and $\omega_2=\omega+\delta\omega_2$ undergo sum-frequency mixing in a quasi-phasematched waveguide, generating an output at $\omega_3 = \omega_1 + \omega_2$. The two-dimensional SFT transfer function is obtained by independently sweeping the two pump detunings, $\delta\omega_1$ and $\delta\omega_2$, and measuring the normalized sum-frequency power. The geometry of the constant-power contours is determined by the waveguide dispersion, whereas longitudinal spatial inhomogeneities modify the associated spectral envelope. The SFT transfer function can be fit to simultaneously extract dispersion orders and a low-order polynomial approximation of the longitudinal phase-mismatch variations.}
    \label{fig:overview}
\end{figure*}

Historically, several diagnostic approaches have been developed for traveling-wave nonlinear optical devices, but existing methods generally involve tradeoffs between throughput, spatial resolution, and sensitivity to second- and higher-order dispersion. The conventional second-harmonic generation (SHG) transfer function, obtained by measuring the second-harmonic conversion efficiency as a function of fundamental wavelength, is a widely used nonlinear diagnostic. This technique can measure the group-velocity mismatch (GVM) between the fundamental and second-harmonic waves and provides a qualitative measure of spatial inhomogeneities~\cite{fejer1992quasi}. Moreover, since the SHG transfer function is determined only by linear combinations of second- and higher-order dispersion terms, it cannot be used to resolve these terms independently. Several advanced techniques have been developed to better resolve spatial inhomogeneities. Complex SHG transfer functions, obtained by measuring the amplitude and phase of the input and output fields, can recover both the longitudinal phase-mismatch profile and longitudinal variations in the nonlinear coupling with high precision, but typically rely on prior knowledge of the waveguide dispersion relations~\cite{chang2014complex}. Additionally, typical implementations use frequency-resolved optical gating to recover the spectral phase of the input and output fields, which limits throughput. More recently, scattered-light imaging has been used to resolve the spatial evolution of the generated second harmonic~\cite{zhao2023unveiling}. When combined with a swept fundamental to obtain a spatially resolved SHG transfer function, this approach also recovers longitudinal variations in both the nonlinear coupling and phase mismatch. While both of these methods can resolve spatial variations in the phase mismatch and nonlinear coupling with micron-scale resolution, they have limited throughput, and neither has been used to recover the dispersion relations of a waveguide.

Three-wave mixing and parametric down-conversion have been relatively unexplored as diagnostic tools for nonlinear waveguides. Although parametric down-conversion is commonly used to measure the dispersion relations of bulk materials~\cite{Jundt1997,Zelmon1997}, extending this technique to guided-wave devices requires the production of many nominally identical waveguides with different quasi-phasematching (QPM) periods. An alternative approach is to replace the varying QPM period with a tunable pump laser~\cite{lenzini2018direct,brecht2011quantum,ansari2018tailoring,xin2022spectrally}. This technique, commonly referred to as stimulated emission tomography~\cite{liscidini2013stimulated}, is typically used for joint-spectral measurements of the generated signal and idler~\cite{triginer2020understanding}. Although stimulated emission tomography contains information about both spatial variations and waveguide dispersion, its use as a general diagnostic for nonlinear waveguides has not been explored.

In this work, we establish three-wave mixing as a rapid diagnostic of dispersion and longitudinal inhomogeneity in nonlinear waveguides. We focus here on the case of sum-frequency tomography (SFT), which we define analogously to stimulated-emission tomography as the transfer function obtained by independently varying the frequency detunings of two input lasers, as shown schematically in Fig.~\ref{fig:overview}. Although the analysis developed here can be extended to any three-wave mixing process, we choose to focus on SFT since this facilitates a direct experimental comparison with traditional SHG-based approaches. In contrast with SHG, SFT produces a two-dimensional transfer function that maps the generated output power as a function of the frequency detunings of two input lasers. A key feature of SFT is that the waveguide dispersion relations determine the geometry of the constant-power contours within this two-dimensional frequency space. Because longitudinal variations weakly perturb the dispersion orders, they primarily modify the spectral envelope of the transfer function without distorting the associated constant-power contours, allowing dispersion and spatial inhomogeneity to be modeled separately. To validate this technique, we characterize two commercially available nonlinear waveguides designed for SHG and optical parametric amplification (OPA), respectively. In both devices, we recover the group-velocity mismatch (GVM) between the interacting waves, the group-velocity dispersion (GVD) of the fundamental wave, and a low-order polynomial approximation of the longitudinal phase-mismatch profile.

\section{Theoretical Framework of Sum-Frequency Tomography}\label{sec:theory}
In this section, we develop the theoretical framework for sum-frequency tomography used throughout this work to characterize nonlinear waveguides. As illustrated in Fig.~\ref{fig:overview}, SFT is performed by pumping a $\chi^{(2)}$ nonlinear waveguide with two tunable CW lasers near degeneracy. We write the pump frequencies as $\omega_1=\omega+\delta\omega_1$ and $\omega_2=\omega+\delta\omega_2$, which together generate a sum-frequency signal at $\omega_3=\omega_1+\omega_2=2\omega+\delta\omega_1+\delta\omega_2$. The measured output SFG power, $P_\text{out}(\delta\omega_1, \delta\omega_2)$, is normalized by the input powers $P_1$ and $P_2$ to give the SFT transfer function, $H(\delta\omega_1, \delta\omega_2)$. In the absence of spatial inhomogeneities, this transfer function takes the familiar $\mathrm{sinc}^2$ form, from which dispersion orders can be extracted by curve fitting. Here, we extend this model to account for spatially inhomogeneous phase-matching conditions and their effects on the SFT transfer function. A key result of this section is that, under the approximations introduced here, the constant-power contours of the SFT transfer function are determined solely by the dispersive contribution to the phase mismatch. In addition, we introduce a compact representation of the waveguide inhomogeneities that modifies the intensity envelope without changing the geometry of these contours. Together, these two features enable the effects of dispersion and device inhomogeneity to be characterized independently, providing a robust framework for parameter extraction.

We model SFT in a nonlinear waveguide using the undepleted-pump approximation, in which the pump powers remain constant throughout the waveguide. In this regime, the coupled-wave equation (CWE) for the generated sum-frequency amplitude becomes
\begin{align}
\partial_z A_3(z)
= -i\kappa A_1(0)A_2(0)
\exp\!\left(i\Delta\phi(\delta\omega_1,\delta\omega_2,z)\right),
\end{align}
where $\kappa$ is the nonlinear coupling coefficient~\cite{jankowski2024ultrafast} and $A_j(z)$ are the complex field amplitudes. The field amplitudes are normalized such that $|A_j(z)|^2$ represents the optical power at frequency $\omega_j$ with $(j\in\{1,2,3\})$. The propagation coordinate is $z\in[0,L]$, where $L$ is the length of the nonlinear section within the waveguide. Here, the function
\begin{align}
\Delta\phi(\delta\omega_1,\delta\omega_2,z)
= \int_0^z \Delta k(\delta\omega_1,\delta\omega_2,z')\,dz'
\end{align}
represents the accumulated phase mismatch associated with the three-wave interaction. We explicitly include the spatial dependence of $\Delta k$ to account for spatial inhomogeneities in the local phase-matching conditions. In general, the nonlinear coupling coefficient $\kappa$ depends on the transverse mode overlap and the local duty cycle of the QPM grating and therefore may vary along the waveguide in addition to $\Delta k$. Throughout this work, we assume that these longitudinal variations in the nonlinear coupling are sufficiently weak that $\kappa$ can be approximated as independent of $z$.

We define the SFT transfer function analogously to the SHG transfer function as the output power $|A_3(L)|^2$ normalized to the corresponding output power obtained for perfect phase matching,
\begin{widetext}
\begin{align}
H(\delta\omega_1,\delta\omega_2) = \frac{|A_3(L)|^2}{\kappa^{2}|A_1(0)|^2 |A_2(0)|^2 L^2}
= \left|\frac{1}{L}\int_0^L \exp\!\bigl(i\,\Delta\phi(\delta\omega_1,\delta\omega_2,z)\bigr)\,dz\right|^2,\label{eqn:tf}
\end{align}
\end{widetext}
where Eqn.~\ref{eqn:tf} follows from approximating $\kappa$ as independent of $z$. This model relates the experimentally measured output power to the accumulated phase mismatch $\Delta\phi$. Eqn.~\ref{eqn:tf} can be used to calculate the SFT transfer function once the launching and collection efficiency of each wave has been characterized, so that the on-chip pump powers, $P_j=|A_j(0)|^2$, and the generated sum-frequency power are known. $\kappa$ is typically measured by varying the pump power to obtain a slope efficiency $\kappa^2$ in $\%/\mathrm{W}-\mathrm{cm}^2$. In this work, we make no attempt to extract $\kappa$ and instead focus on the shape of the transfer function by normalizing the peak value to 1, which is sufficient to characterize the spatial inhomogeneities of the waveguide and the dispersion relations of the interacting waves.

\begin{figure}[htbp]
    \centering
    \includegraphics[width=0.9\linewidth]{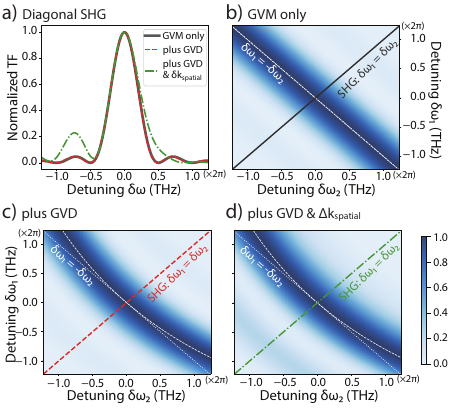}
    \caption{Comparison of the SHG and SFT transfer function in the presence of spatial inhomogeneity and dispersion. (a) SHG transfer functions for three regimes: pure GVM (black solid), GVM with added GVD (red dashed), and GVM with added GVD plus a spatial-inhomogeneity term (green dash-dotted). (b)--(d) Two-dimensional SFT transfer functions for the same three regimes; the corresponding SHG transfer functions are given by the diagonal slices $\delta\omega_1 = \delta\omega_2$. The dashed white line indicates the phase-matching contour $\Delta k_\mathrm{disp} (\delta\omega_1,\delta\omega_2) + \Delta k_0=0$. (b) Without GVD, the transfer function is phase-matched along $\delta \omega_1 = -\delta \omega_2$. (c) GVD bends the phase-matching contour away from the anti-diagonal. (d) Spatial inhomogeneities modify the intensity envelope without changing the contour geometry.}
    \label{fig:transferfun}
\end{figure}

Using this model, we can formulate our goal as an inverse problem: inferring the waveguide parameters that determine the accumulated phase mismatch $\Delta\phi(\delta\omega_1,\delta\omega_2,z)$ from the experimentally measured transfer function $H(\delta\omega_1,\delta\omega_2)$. Once $\Delta \phi(\delta \omega_1,\delta\omega_2,z)$ is obtained, the local phase mismatch can be calculated using $\Delta k(\delta \omega_1,\delta\omega_2,z) = \partial_z \Delta \phi(\delta \omega_1,\delta\omega_2,z)$. In many cases, contributions to the phase mismatch are dominated by a single geometric parameter, such as variations in film thickness, $\Delta h(z)$, in which case, to first order, perturbations to the local phase mismatch can be related to local spatial variations by $\Delta k(z) = (\partial\Delta k/\partial h)\Delta h(z)$. For typical waveguides, spatial variations weakly perturb the dispersive contributions to the phase mismatch, allowing us to write the accumulated phase mismatch as
\begin{align}
\Delta\phi(\delta\omega_1,\delta\omega_2,z) =\Delta\phi_\text{spatial}(z)+\Delta k_\text{disp}(\delta\omega_1,\delta\omega_2)\cdot z.~\label{eqn:separable_pm}
\end{align}
This approximation is valid when longitudinal variations produce negligible changes to the dominant dispersion orders, namely, the group-velocity mismatch between the interacting waves and the group-velocity dispersion of each wave. An immediate consequence of Eqn.~\ref{eqn:separable_pm} is that the contours of constant $\Delta k_\text{disp}(\delta\omega_1,\delta\omega_2)$ are also contours of constant $H(\delta\omega_1,\delta\omega_2)$. This property allows the dispersive contribution to the phase mismatch to be characterized independently of the spatial variations of the waveguide.

The first term, $\Delta\phi_\text{spatial}(z)$, accounts for frequency-independent contributions to the accumulated phase mismatch arising from an overall error in the waveguide geometry or from spatial inhomogeneities. Physically, the derivative of this term, $\partial_z\Delta\phi_\text{spatial}(z)$, represents the local phase mismatch of the SHG process ($\omega+\omega\rightarrow2\omega$) at position $z$. To concisely parameterize the spatial dependence of $\Delta\phi_\text{spatial}(z)$, we series-expand the accumulated phase mismatch as
\begin{align}
\Delta\phi_{\mathrm{spatial}}(z)
=
\sum_{m=1}^{\infty} c_m \mathcal{P}_m(x),
\quad
x=\frac{2z}{L}-1,
\end{align}
where we omit the $m=0$ term because it contributes only a frequency- and position-independent phase to the integrand in Eqn.~\ref{eqn:tf} and therefore has no effect on the measured transfer function. Although any orthogonal basis $\mathcal P_m(x)$ on the interval $x\in[-1,1]$ could be used in principle, we choose the Legendre polynomials, $\mathcal{P}_1(x)=x$, $\mathcal{P}_2(x)=\frac{1}{2}(3x^2-1)$, $\mathcal{P}_3(x)=\frac{1}{2}(5x^3-3x)$, \dots, which exhibit a number of useful properties for calculating SFT transfer functions. First, we note that this polynomial series expansion for the accumulated phase mismatch can easily be converted into more commonly encountered descriptions of spatial inhomogeneities, such as a Taylor series expansion of the local phase mismatch. The lowest-order term, $\mathcal{P}_1(x)$, corresponds to the spatially constant local phase mismatch, $\Delta k_0 = 2 c_1/L$, and when combined with $\Delta k_\text{disp}$ this term determines the phase-matched wavelengths. The higher-order terms parameterize spatial variations about this constant phase mismatch. Second, since each term in this expansion has zero mean, $\int_{-1}^{1} \mathcal{P}_m(x) dx = 0$, the higher-order terms do not shift the peak position to first order in their coefficients. This property follows from a perturbative expansion of the integral in Eqn.~\ref{eqn:tf} around the peak of the transfer function ($c_1 = 0$) assuming the higher-order coefficients are small. Numerically, we find that the resulting peak shift remains negligible when the higher-order coefficients satisfy $|c_m|\ll\pi$. Finally, we note that this series is observed to converge quickly for typical waveguides, since higher-order contributions tend to introduce increasingly rapid oscillations to $\Delta \phi_\mathrm{spatial}(z)$ that are averaged out by the integral in Eqn.~\ref{eqn:tf}. This motivates truncating the series at low $m$, and in this work we retain terms through $m=3$.

\begin{figure*}[t]
    \centering
    \includegraphics[width=0.8\linewidth]{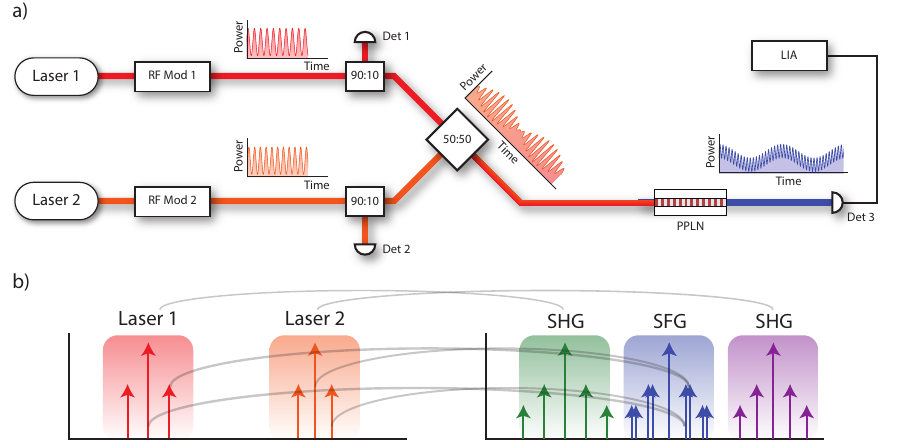}
    \caption{Experimental apparatus for SFT. (a) Optical setup. Laser 1 (Santec TSL-570) and Laser 2 (Santec TSL-770) are continuous-wave tunable lasers; each is driven through an electro-optic modulator (RF Mod) at a frequency $\Omega_1$ and $\Omega_2$, respectively. Polarization-maintaining (PM) 90:10 fiber tap couplers route a small fraction of each pump to InGaAs power detectors (Det 1, 2) for input-power normalization, while the main beams are combined in a PM 50:50 fiber coupler and launched into the PPLN waveguide. The generated light is detected on a silicon photodiode (Det 3) and demodulated by a lock-in amplifier (LIA). The time-domain insets show the modulated power on each arm and the resulting beat pattern after combination. (b) Frequency-domain picture of the nonlinear output. The modulated lasers near 1560~nm (red and orange) generate three signals: SHG at  $2\omega_1$ and $2\omega_2$ (green and purple), and the desired SFG at $\omega_1 + \omega_2$ (blue). Interference between the upconverted sidebands generates a beatnote, $\Omega_1 \pm \Omega_2$, unique to the SFG signal, which is isolated by lock-in detection.}
    \label{fig:apparatus}
\end{figure*}

The second term, $\Delta k_\text{disp}(\delta\omega_1,\delta\omega_2)\cdot z$, accounts for the frequency dependence of the phase mismatch due to the dispersion relations of the waveguide. Since the frequency-independent contributions are included in $\Delta\phi_{\mathrm{spatial}}(z)$, the dispersive contributions to the phase mismatch, $\Delta k_\text{disp}(\delta\omega_1,\delta\omega_2)$, vanish by construction at the degeneracy point $\delta\omega_1=\delta\omega_2=0$. Taylor expanding around this point, we obtain
\begin{align}
\Delta k_\text{disp}(\delta\omega_1,\delta\omega_2) &= \Delta k'\,(\delta\omega_1+\delta\omega_2)
 - \tfrac{1}{2}k_1''\,(\delta\omega_1^2+\delta\omega_2^2) \notag\\
 &\quad + \tfrac{1}{2}k_3''\,(\delta\omega_1+\delta\omega_2)^2 + \mathcal{O}(\delta\omega_j^3),
\end{align}
where the first-order coefficient $\Delta k'\equiv v_{g,3}^{-1}-v_{g,1}^{-1}=v_{g,3}^{-1}-v_{g,2}^{-1}$ is the group-velocity mismatch (GVM) between the sum-frequency and fundamental waves, and the equality follows from the degeneracy of the two fundamental waves. The second-order contributions, $k_1''=k_2''$ and $k_3''$, represent the group-velocity dispersion (GVD) evaluated at $\omega$ and $2\omega$, respectively. For the waveguides studied here, we truncate this expansion at second order in $\delta \omega_j$, since these leading-order terms dominate over higher-order contributions across the experimentally accessible range of $\delta\omega_1$ and $\delta \omega_2$. Higher-order dispersion terms become increasingly important in waveguides engineered to have $\Delta k'\approx 0$~\cite{jankowski2020ultrabroadband}. A more detailed analysis of which dispersion orders can be neglected and, conversely, which dispersion orders can be extracted using SFT is given later in Sec.~\ref{sec:sensitivity}. 

Figure~\ref{fig:transferfun} illustrates the mapping between different contributions to $\Delta \phi$ and features of the SHG and SFT transfer functions. We first show numerically simulated SHG transfer functions in Fig.~\ref{fig:transferfun}(a) for three cases: (i) an idealized waveguide with GVM but no GVD or spatial inhomogeneities, (ii) a waveguide with $k_1''$, but no spatial inhomogeneities, and (iii) a waveguide with both $k_1''$ and spatial inhomogeneities. In the absence of spatial inhomogeneities, the transfer function takes the form $\mathrm{sinc}^2(\Delta k L/2)$, with a characteristic width determined by the GVM between the fundamental and second harmonic waves. For the parameters considered here, the addition of $k_1''$ produces little discernible change in the SHG transfer function, consistent with the limited sensitivity of SHG to the individual dispersion orders. Adding spatially inhomogeneous phase mismatch distorts the transfer function, in this case producing an asymmetric side lobe, which provides a qualitative indication of longitudinal phase-matching homogeneity but makes extraction of the GVM less reliable. As shown in Fig.~\ref{fig:transferfun}(b) and (c), the two-dimensional SFT transfer function reveals the effect of $k_1''$  as curvature of the phase-matching contour about the $\delta \omega_1 = -\delta \omega_2$ anti-diagonal. Adding spatial inhomogeneities (Fig.~\ref{fig:transferfun}(d)) modifies the spectral envelope without changing this contour geometry, since $k_1''$ has been held constant. We emphasize here that while this behavior allows dispersive contributions to $\Delta\phi$ to be characterized independently of spatial variations, it does not imply that every dispersion order can be extracted.

\section{Experimental Characterization of PPLN Waveguides}\label{sec:exp}

In this section, we experimentally validate SFT using two nominally identical PPLN waveguides purchased from HC Photonics. Since the dispersion relations of these weakly confining waveguides are only weakly perturbed by the variations in waveguide geometry, we expect to recover similar values for the dispersion orders of each waveguide despite them having different longitudinal variations. Comparing the parameters extracted from the two devices therefore provides an internal consistency check on the parameters retrieved by SFT, while differences between the devices can reveal systematic effects or limitations of the method. Both waveguides are \SI{15}{mm} long, with nominally identical QPM periods chosen to phase-match a degenerate three-wave mixing process, $\SI{1560}{nm} + \SI{1560}{nm} \leftrightarrow \SI{780}{nm}$. The first module is a dedicated second-harmonic generation module (HCP-SHG) designed to frequency-double a \SI{1560}{nm} pump to \SI{780}{nm}, and the second is a degenerate optical parametric amplification module (HCP-OPA) designed to amplify a \SI{1560}{nm} signal with a \SI{780}{nm} pump.

Figure~\ref{fig:apparatus}(a) shows the experimental setup used for SFT. We employ two tunable CW lasers, hereafter referred to as Laser 1 (Santec TSL-570) and Laser 2 (Santec TSL-770), to pump the PPLN waveguide. The optical frequency of each laser is denoted by $\omega_1$ and $\omega_2$, respectively, and their wavelengths are independently varied from \SI{1500}{nm} to \SI{1620}{nm}. Laser 1 is step tuned to coarsely sample slices of the SFT transfer function, while laser 2 is swept continuously across the wavelength range. Since each continuous sweep of Laser 2 takes approximately \SI{0.6}{s}, and we find that only 5 - 10 sweeps are sufficient for SFT to recover the relevant dispersion orders, a single SFT transfer function can be acquired in less than 10 seconds. Each laser is modulated using an electro-optic intensity modulator, after which the two beams are combined using a PM 50:50 fiber coupler and launched into the PPLN waveguide. The generated optical output is detected by a photoreceiver, with the resulting output voltage processed using a lock-in amplifier. Finally, the measured SFG signal is normalized by the powers of the pump lasers, which are tapped off and measured before being sent to the waveguide (Det 1 and 2 in Fig.~\ref{fig:apparatus}). Because the absolute optical coupling and detection efficiencies are not calibrated, this experimentally measured transfer function differs from the theoretical transfer function by an overall frequency-independent prefactor. All beam routing from the lasers to the PPLN waveguide is performed with polarization-maintaining 1560-nm fiber, while single-mode 780-nm fiber is used to route the output signal from the PPLN module to the detector.

Importantly, in addition to the desired SFG signal at frequency $\omega_1+\omega_2$, the PPLN waveguide also produces undesired SHG signals at $2\omega_1$ and $2\omega_2$, all of which contribute to the detected signal. Since the frequencies of the three output signals are too close to be separated using optical filters, we instead use lock-in detection to selectively extract the SFG signal. We achieve this by using electro-optic intensity modulators to attach distinct radiofrequency (RF) labels at frequencies $\Omega_1$ and $\Omega_2$ to the two pump lasers, as shown in Fig.~\ref{fig:apparatus}(b). The power in each optical signal can be measured by demodulating the photoreceiver output at the corresponding RF frequency, up to an overall constant. These modulations on the pump intensities translate into modulations of the SFG signal at $\Omega_1\pm \Omega_2$, whereas the SHG processes produce only harmonics of the individual modulation frequencies, at integer multiples of $\Omega_1$ or $\Omega_2$. Therefore, by demodulating the photoreceiver signal at the reference frequency of $|\Omega_1-\Omega_2|$ with a lock-in amplifier, we can selectively isolate the SFG contribution near degeneracy.

\begin{figure}[htbp]
    \centering
    \includegraphics[width=0.9\linewidth]{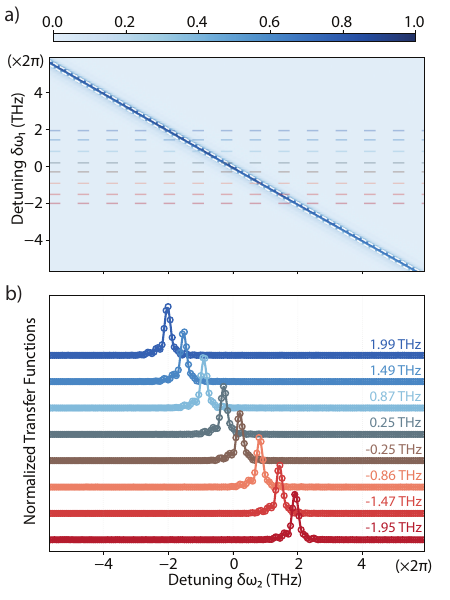}
    \caption{SFT measurement and best-fit model for the HCP-SHG waveguide. The frequency detunings, $\delta\omega_1$ and $\delta\omega_2$, are reported relative to a reference wavelength of 1560 nm. (a) Measured normalized transfer function $H(\delta\omega_1,\delta\omega_2)$ across the full scanned detuning range. The white dashed contour marks the half-maximum level of the best-fit model, and the colored horizontal dashed lines indicate the values of $\delta\omega_1$ plotted below. (b) Waterfall plot of the corresponding slices, vertically offset for clarity. Open circles show the measured data, and solid curves show the best-fit SFT model.}
    \label{fig:results}
\end{figure}

In Fig.~\ref{fig:results}(a), we show the experimentally measured SFT transfer function for the HCP-SHG waveguide. Because small fluctuations occur when the laser changes sweep direction, the data near the corners contain minor artifacts. Although the experimental scan covered \SI{1500}{nm}--\SI{1620}{nm}, only the \SI{1507}{nm}--\SI{1613}{nm} range was used for fitting to avoid these artifacts. We fit this data using Eqn.~\ref{eqn:tf}, including a constant prefactor, to extract the dispersion orders ($\Delta k'$, $k_1''$, $k_3''$) and the Legendre coefficients ($c_1$, $c_2$, $c_3$). The 3-dB contour of the fitted SFT transfer function is overlaid as dashed white lines in Fig.~\ref{fig:results}(a), showing strong agreement with the experimental data. Fig.~\ref{fig:results}(b) compares slices of the SFT transfer function (open circles) with the fitted transfer function (solid lines), showing excellent agreement between the experimentally measured transfer function and our model. Together, these plots suggest that the model captures the observed behavior of the waveguide with the included dispersion orders and Legendre coefficients. The same procedure is applied to the HCP-OPA waveguide, and the parameters estimated for the two devices are summarized in Table~\ref{tab:dispersion_params1}.

\begin{table*}[t]
    \centering
    \caption{Dispersion orders and spatial-inhomogeneity parameters extracted from the SFT transfer function of the HCP-SHG and HCP-OPA waveguides. The extracted values are compared with simulated values obtained using a finite-difference eigensolver~\cite{Fallahkhair2008}.}
    \label{tab:dispersion_params1}
    \begin{tabular}{lccc}
        \toprule
        \textbf{Parameter} & \textbf{HCP-SHG} & \textbf{HCP-OPA} & \textbf{Simulation} \\
        \midrule
        Phase mismatch $\Delta k_0$ (mm$^{-1}$)           & $-0.035$ & $-0.082$ & --     \\
        GVM $\Delta k'$ (fs/mm)                           & $279.2$   & $267.1$   & $272.8$  \\
        Fundamental GVD $k_1''$ (fs$^2$/mm)               & $60.9$    & $50.7$    & $59.3$   \\
        Sum-frequency GVD $k_3''$ (fs$^2$/mm)             & $2261.4$  & $757.1$   & $370.0$  \\
        Legendre coefficient $c_2$ (rad)                    & $1.10$  & $0.26$  & --     \\
        Legendre coefficient $c_3$ (rad)                    & $0.27$ & $-0.43$ & --     \\
        \bottomrule
    \end{tabular}
\end{table*}

We reiterate here that for the two waveguides we expect only the dispersion orders to agree, since these parameters are weakly dependent on waveguide geometry. In practice, the parameters listed in Table~\ref{tab:dispersion_params1} show varying levels of agreement. For cross-validation, we numerically estimate the waveguide dispersion using the nominal waveguide geometry~\cite{cho2021power}, with the refractive index of the guided region calculated using~\cite{umemura2014sellmeier,umemura2016thermo}. The group-velocity mismatch $\Delta k'$ shows good agreement between both waveguides and the simulated value, suggesting that this parameter can be reliably extracted by SFT. Similar agreement is observed for the fundamental GVD, $k_1''$, which again suggests this parameter can be reliably extracted by SFT. In contrast, the sum-frequency GVD, $k_3''$, exhibits a large discrepancy between the two waveguides and the simulated value. Since the core diameter of the waveguide is much larger than the sum-frequency wavelength of \SI{780}{nm}, the propagation constant of this mode should be close to that of the bulk medium, consistent with the simulated value. Therefore, waveguide-to-waveguide fabrication variations are unlikely to explain the discrepancy in $k_3''$; instead, the large variation in the extracted values suggests that $k_3''$ is only weakly determined by the measured SFT transfer function.

\begin{figure}[htbp]
    \centering
    \includegraphics[width=0.9\linewidth]{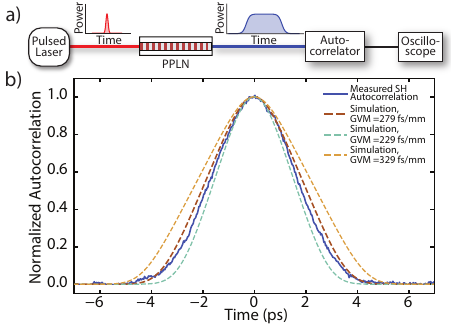}
    \caption{Validation of the SFT-extracted GVM via second-harmonic autocorrelation. (a) Experimental setup: mode-locked transform-limited 1.14~ps sech$^2$ pulses centered at 1558.4~nm are coupled into the 15~mm HCP-SHG PPLN waveguide, and the generated 779.2~nm second-harmonic pulses are characterized using a non-collinear, background-free intensity autocorrelator. (b) Measured autocorrelation trace (solid blue) compared with single-pass SHG simulations for three values of the group-velocity mismatch, $\Delta k' \in  \lbrace 279, 229, 329\rbrace $~fs/mm (brown, cyan, and orange dashed, respectively).}

    \label{fig:autocorrelation}
\end{figure}

We further validate the extracted GVM by driving the HCP-SHG device with picosecond pulses and measuring the intensity autocorrelation of the generated second-harmonic pulse, as shown in Fig.~\ref{fig:autocorrelation}. Mode-locked, transform-limited \SI{1.14}{ps} pulses centered at \SI{1558.4}{nm}, with a sech$^2$ spectral envelope, are generated by an ultrafast optical clock (PriTel UOC-5-20G) operating at a repetition rate of \SI{10}{GHz}. These pulses are coupled into the waveguide, producing frequency-doubled pulses centered at \SI{779.2}{nm}, which are characterized using a non-collinear, background-free intensity autocorrelator (Femtochrome FR-103XL). We first measured the pulse duration of the fundamental using the autocorrelator and found that our input pulses have a transform-limited $\text{sech}^2(t/\tau_p)$ power envelope, where a characteristic pulse width of $\tau_p = \SI{0.65}{ps}$ corresponds to an intensity FWHM of \SI{1.14}{ps}. In waveguides where GVM is the dominant dispersion order and the fundamental remains undepleted, the time-domain power envelope of a second-harmonic pulse generated by SHG of a $\text{sech}^2$ fundamental is given by
\begin{align}
I_{2\omega}(t) \propto \left| \frac{\tau_p}{\Delta k'} \left[ \tanh \left( \frac{t}{\tau_p} \right) - \tanh \left( \frac{t - \Delta k'L}{\tau_p} \right) \right] \right|^2,
\end{align}
which is effectively a top-hat pulse with apodized edges~\cite{jankowski2024ultrafast}. The temporal walk-off, $\Delta k'L$, sets the duration of the resulting top-hat-like shape. The intensity autocorrelation of the second-harmonic pulse is given by
\begin{align}
G^{(2)}(\tau)= \int_{-\infty}^{\infty} I_{2\omega}(t) I_{2\omega}(t - \tau)\, dt.
\end{align}
In the limit of large temporal walk-off, the generated second harmonic therefore produces a triangle-like autocorrelation trace, with a characteristic width set by the temporal walk-off of the waveguide.

For intensity-autocorrelation measurements, it is generally desirable to keep the SHG conversion efficiency low to avoid distortion of the pulse shape. In our measurement, however, the pump power was increased to ensure a sufficient signal-to-noise ratio (SNR), resulting in a 5\% conversion efficiency (\SI{0.6}{mW} of SHG power for \SI{12}{mW} of pump power). Although the level of pump depletion is relatively small, it can still distort the second-harmonic pulse into a temporally decaying profile and result in an artificial narrowing of the intensity autocorrelation trace. As shown in Fig.~5(b), the measured autocorrelation profile has tails consistent with the \SI{279}{fs/mm} GVM inferred from SFT, but is narrower near its half-maximum, which we attribute to the aforementioned artificial narrowing caused by pump depletion. We also show theoretical autocorrelations calculated for GVM values perturbed by $\pm\SI{50}{fs/mm}$, with the resulting traces showing visible disagreement with the experimental data. Together, these results provide further independent validation of the GVM extracted by SFT. We note that autocorrelation measurements of the temporal walk-off involve a trade-off between SNR and pump depletion, limiting the precision with which the GVM can be extracted from the autocorrelation trace.

\section{Parameter Sensitivities and Diagnostic Limits}\label{sec:sensitivity}
\begin{figure*}[t]
    \centering
    \includegraphics[width=0.9\linewidth]{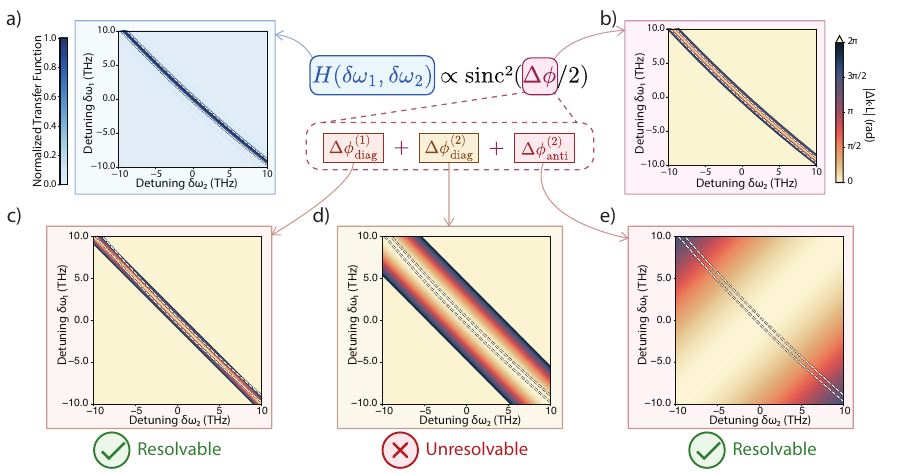}
    \caption{Geometric decomposition of the SFT transfer function and dispersion order resolvability. The central expression $H(\delta\omega_1,\delta\omega_2)\propto\mathrm{sinc}^2(\Delta\phi/2)$ connects the measured transfer function to the accumulated phase mismatch, which is decomposed as $\Delta\phi=\Delta\phi_{\mathrm{diag}}^{(1)}+\Delta\phi_{\mathrm{diag}}^{(2)}+\Delta\phi_{\mathrm{anti}}^{(2)}$. (a)~Simulated normalized transfer function $H(\delta\omega_1,\delta\omega_2)$. (b)~Total accumulated phase mismatch $|\Delta\phi|$, capped at $2\pi$. (c)--(e)~Magnitudes of the individual contributions to $\Delta\phi$ plotted on the same color scale: (c)~the first-order diagonal term $\Delta\phi_{\mathrm{diag}}^{(1)}=\Delta k'(\delta\omega_1+\delta\omega_2)L$; (d)~the second-order diagonal term $\Delta\phi_{\mathrm{diag}}^{(2)} =\frac{1}{4}(2k_3''-k_1'')(\delta\omega_1+\delta\omega_2)^2L$; and (e)~the second-order anti-diagonal term $\Delta\phi_{\mathrm{anti}}^{(2)}=-\frac{1}{4}k_1''(\delta\omega_1-\delta\omega_2)^2L$. White dashed contours in (b)--(e) indicate the half-maximum contour, $H=0.5$, of the simulated SFT transfer function and therefore outline the region in which the transfer-function signal remains appreciable. The phase contributions in (b)--(e) are plotted on a common color scale from $0$ to $2\pi$, with values exceeding $2\pi$ shown using the over-range color. A dispersion contribution is resolvable when it produces appreciable phase variation within the high-signal region enclosed by the half-maximum contour. In contrast, a contribution that remains small throughout this region and becomes significant only where the transfer function is already suppressed has little observable effect on the measured SFT transfer function and is therefore difficult to resolve.}
    
    \label{fig:contribution}
\end{figure*}

The experimental results in the previous sections showed reasonable agreement for both $\Delta k'$ and $k_1''$, but the strong disagreement in the values extracted for $k_3''$ suggests that SFT is not sufficiently sensitive to this parameter. In this section, we provide an intuitive physical picture of how each dispersion order is mapped into the shape of the SFT transfer function and discuss which waveguide parameters can be extracted more reliably. For simplicity, we consider the case with no longitudinal spatial inhomogeneity. In this limit, the SFT transfer function reduces to a simplified form
\begin{align}
    H(\delta \omega_1,\delta \omega_2)\propto \mathrm{sinc}^2\left(\frac{1}{2}\Delta \phi\right),
\end{align}
where $\Delta \phi$ is the accumulated phase mismatch. The structure of $H$ determines which contributions to $\Delta \phi$ can be resolved in an experiment. We show the structure of $H$ and $\Delta \phi$ in Fig.~\ref{fig:contribution}(a) and (b), respectively, for a set of representative waveguide parameters.
Note that the magnitude of the SFT transfer function drops rapidly in the region where $|\Delta \phi|$ is large, with $H\propto (\Delta \phi)^{-2}$ for large values of $|\Delta\phi|$. For the sensitivity analysis below, we therefore consider the central sinc lobe, $|\Delta \phi|<2\pi$, as the region over which the SFT transfer function provides the most reliable information.

The accumulated phase mismatch consists of contributions from each of the three dispersion orders considered here, $\Delta \phi = \Delta \phi_{\Delta k'}+\Delta \phi_{k_1''}+\Delta \phi_{k_3''}$, with contributions from the GVM given by $\Delta \phi_{\Delta k'}=\Delta k'(\delta \omega_1+\delta \omega_2)L$, and the contributions from the GVD of the fundamental and sum-frequency given by $\Delta \phi_{k_1''}=-\frac{1}{2}k_1''(\delta \omega_1^2+\delta \omega_2^2)L$, and $\Delta \phi_{k_3''}=\frac{1}{2}k_3''(\delta \omega_1+\delta \omega_2)^2L$, respectively. These contributions predominantly vary along the diagonals in the $\omega_1-\omega_2$ plane, and therefore we remap these terms explicitly into diagonal and anti-diagonal contributions as
\begin{widetext}
\begin{align}
    \Delta \phi=\underbrace{\Delta k'(\delta \omega_1+\delta \omega_2)L}_{\mathrm{1^{st}~order~diagonal}:~\Delta\phi_{\mathrm{diag}}^{(1)}}+\underbrace{\frac{1}{4}\left(2k_3''-k_1''\right)(\delta \omega_1+\delta \omega_2)^2L}_{\mathrm{2^{nd}~order~diagonal}:~\Delta\phi_{\mathrm{diag}}^{(2)}} -\underbrace{\frac{1}{4}k_1''(\delta \omega_1-\delta\omega_2)^2L}_{\mathrm{2^{nd}~order~anti~diagonal}:~\Delta\phi_{\mathrm{anti}}^{(2)}}.
\end{align}
\end{widetext}
The first- and second-order terms, $\Delta\phi_{\mathrm{diag}}^{(1)}$ and $\Delta\phi_{\mathrm{diag}}^{(2)}$, respectively, depend only on $(\delta \omega_1+\delta\omega_2)$, and therefore their contributions to $\Delta \phi$ vary along the diagonal $\delta\omega_1 = \delta\omega_2$. Features along this diagonal therefore carry information about the GVM $\Delta k'$ and a weighted sum of the GVDs, $(2k_3''-k_1'')$. The second-order diagonal contribution $\Delta\phi_{\mathrm{diag}}^{(2)}$ becomes significant only for sufficiently large $(\delta \omega_1+\delta\omega_2)$. If this term only becomes significant outside the central lobe, where $|\Delta\phi|\gg2\pi$, it has little effect on the experimentally measurable transfer function since the magnitude of $H$ is strongly suppressed in this region. To quantify the relative importance of the first-order and second-order diagonal terms within the central lobe of the sinc function, we evaluate the second-order contribution at the detuning where the first-order contribution reaches $2\pi$, $(\delta \omega_1+\delta\omega_2)=\frac{2\pi}{\Delta k' L}$. The second-order contribution at this detuning is $\frac{\pi^2(2k_3''-k_1'')}{(\Delta k')^2 L}$, which is only $\SI{6e-3}{}$ for the nominal simulated waveguide parameters (see Table~\ref{tab:dispersion_params1}). Therefore, compared with the first-order term (i.e., GVM), the second-order term is too small to produce appreciable variations in the SFT transfer function within the central lobe. Since the second-order diagonal term is the only contribution through which $k_3''$ enters into the accumulated phase mismatch, we cannot reliably extract the value of $k_3''$ from the structure of the SFT transfer function, consistent with the large discrepancies in the values of $k_3''$ extracted from our devices. In contrast, $\Delta k'$ can be reliably estimated since $\Delta\phi_{\mathrm{diag}}^{(1)}$ determines the spectral width of the SFT transfer function along the diagonal. In principle, $k_3''$ can be resolved in waveguides where $\Delta k'$ is engineered to be small.

The fundamental GVD, $k_1''$, can be reliably extracted from the SFT transfer function via the second-order anti-diagonal term, $\Delta\phi_{\mathrm{anti}}^{(2)}$. To see this more clearly, we note that the $\Delta \phi=0$ contour corresponds to the peak of the SFT transfer function in the $(\delta \omega_1, \delta \omega_2)$-plane. Neglecting the second-order diagonal term as insignificant, this contour reduces to
\begin{align}
    (\delta \omega_1+\delta\omega_2)=\frac{k_1''}{4\Delta k'}(\delta \omega_1-\delta\omega_2)^2,
\end{align}
which defines a quadratic curve. Importantly, the curvature of the $\Delta \phi=0$ contour is proportional to $k_1''$, providing a direct means of extracting the fundamental GVD from the maximum-power contour of the SFT transfer function. Putting this together, we find that the parameters we extract reliably throughout this work, $\Delta k'$ and $k_1''$, are closely tied to the width and curvature of the central lobe, respectively. In principle, higher-order dispersive terms can be extracted from features outside the central lobe region $|\Delta\phi|\leq 2\pi$, provided that the SFT transfer function can be measured with sufficiently high signal-to-noise ratio. In this regime, contributions from higher-order dispersion scale more rapidly with detuning and can become comparable to, or even exceed, the lower-order contributions, making them potentially resolvable. Extending SFT beyond second-order dispersion, however, lies outside the scope of this work.

\section{Conclusion}
We have demonstrated sum-frequency tomography (SFT) as a scalable, intensity-only diagnostic for traveling-wave $\chi^{(2)}$ nonlinear waveguides. By extending conventional one-dimensional wavelength-swept SHG to two-dimensional near-degenerate SFG, SFT exposes geometric features that are otherwise hidden: the dispersion relation determines the orientation and curvature of the phase-matching contours of the transfer function, while longitudinal fabrication errors distort its intensity envelope. This multidimensional measurement enables the extraction of dispersion and low-order spatial inhomogeneity without phase-sensitive detection or spatially resolved imaging. Applying SFT to two commercial PPLN modules, we recovered the group-velocity mismatch and the fundamental GVD, obtaining values consistent with waveguide-mode simulations. The extracted GVM was further confirmed by an independent pulsed SHG autocorrelation measurement, providing an experimental cross-check of our scheme. The same measurements also revealed device-dependent longitudinal phase-mismatch variations, demonstrating that SFT can diagnose both dispersion and fabrication-induced spatial inhomogeneity in practical quasi-phasematched waveguides.

An important limitation of our current demonstration is that our scheme cannot extract dispersion orders that do not appreciably affect the structure of the transfer function within the regions where SFG is well phase matched. For the waveguides we tested experimentally, we could extract the GVM $\Delta k'$ and the fundamental GVD $k_1''$, while the sum-frequency GVD $k_3''$ could not be reliably estimated. This is because the contribution of $k_3''$ is masked by that of $\Delta k'$ within the main lobe of the transfer function. A key second limitation of this approach is that we only recover a low-order polynomial expansion for the accumulated phase-mismatch. While this provides an overall measure of waveguide uniformity, it does not recover spatial variations with the same fidelity as the techniques introduced in~\cite{chang2014complex} and~\cite{zhao2023unveiling}.

There are several possible future directions for overcoming these limitations. In principle, lock-in detection recovers both the amplitude and phase of the modulation, providing access to the group delay of the complex transfer function. SFT can therefore be extended to measure complex transfer functions, which could enable more complete recovery of both the dispersion relations and the spatial variations at the cost of greater complexity. An alternative approach is to actively control the phase-matching condition to directly extend the range of pump-wavelength combinations over which SFG occurs. Such control could be realized using fast temperature tuning~\cite{hu2026efficient-adaptive-PM} or programmable QPM gratings~\cite{yanagimoto2026programmable}.

Because SFT uses only swept CW lasers and direct detection, it is well suited for high-throughput device screening, fabrication feedback, and troubleshooting of quasi-phasematched photonic circuits. The approach is not limited to the PPLN modules studied here but can be applied to a wide range of nonlinear photonics platforms, including programmable nonlinear waveguides~\cite{yanagimoto2026programmable} and induced-$\chi^{(2)}$ systems~\cite{timurdogan2017electric, hickstein2019self, billat2017large}. SFT could therefore become a practical in situ diagnostic tool for the next generation of nonlinear photonic systems.

\begin{acknowledgments}
The authors wish to thank NTT Research for their financial and technical support. We thank Martin Fejer, Jeffrey Moses, Yiqi Zhao, Benjamin Ash, Fan Wu, Yanzhou Wu and Mandar Sohoni for helpful comments and discussions.

Y.L., E.N., T.P.M., and M.J. conceived the project. Y.L. and M.J. designed the sum-frequency tomography experiments. Y.L., K.C., and M.J. performed the sum-frequency tomography experiments. Y.L. and K.C. developed the sum-frequency tomography fitting code with contributions from R.Y. and M.J. K.K. performed the autocorrelation measurements and associated simulations. Y.L. developed the waveguide mode simulations with contributions from R.Y. Y.L. analyzed the results and prepared the figures. Y.L., R.Y., and M.J. wrote the manuscript with input from all authors. M.J. supervised the project.
\end{acknowledgments}

\bibliography{ref}

\end{document}